\documentclass[12pt]{article}
\usepackage{amsmath,amssymb}
\usepackage{bbm,latexsym}
\usepackage{epsfig}
\usepackage[numbers,sort&compress]{natbib}
\newcommand{\zz}{\mathbbm{Z}}
\newcommand{\gr}{\mathcal{G}}
\newcommand{\bone}{\mathbbm{1}}
\newtheorem{theorem}{Theorem}
\newtheorem{lemma}{Lemma}
\renewcommand{\mod}[1]{\,(\mathrm{mod}\,{#1})}
\begin{document}
\title{\normalsize \hfill UWThPh-2013-24 \\[1cm] \LARGE
On the characterization of the $SU(3)$-subgroups \\ of type~C and~D}
\author{W. Grimus\thanks{E-mail: walter.grimus@univie.ac.at} 
\setcounter{footnote}{6}
and P.O. Ludl\thanks{E-mail: patrick.ludl@univie.ac.at} \\[4mm]
\small University of Vienna, Faculty of Physics \\
\small Boltzmanngasse 5, A--1090 Vienna, Austria \\[4.6mm]}

\date{December 19, 2013}

\maketitle

\begin{abstract}
We investigate the two classes of finite subgroups of $SU(3)$ that are called
type~C and~D in the book of Miller, Blichfeldt and Dickson. We present two
theorems which fully determine the form of the generators in a suitable basis.
After exploring further properties of these groups, we are able 
to construct a complete list of infinite series in which
these groups are arranged. For type~C there are infinitely many series
whereas for type~D there are only two.
Explicit examples of these series are presented which illustrate the
general results. 
\end{abstract}

\newpage

\section{Introduction}
Finite subgroups of $SU(3)$ are very popular as flavour symmetry
groups in particle physics~\cite{ramond,ishimori,review,ludl-SU3}. 
This is connected to the fact that according to our 
knowledge three families of fermions exist in nature. 
In recent years such flavour groups have been applied, in particular,
in the lepton sector---for reviews see for instance~\cite{altarelli,king}.
The basic classification of all finite subgroups of $SU(3)$ was
performed nearly 100 years ago by Blichfeldt in~\cite{miller}. 
Much later, also particle physicists have focused on such subgroups
and have subjected them to mathematical scrutiny for physics
purposes~\cite{fairbairn,luhn168,ludl,zwicky,principal,comments,merle}. 
Moreover, general properties of finite groups have recently been
reviewed in~\cite{ramond,ishimori,review}. 
Another approach to finite groups is via computer-algebraic
methods. For example, using GAP~\cite{GAP} together with the library
of small groups~\cite{SGL}, 
in~\cite{parattu} \emph{all} finite groups up to order 100 have been
studied with respect to their properties as flavour groups in the
lepton sector.

Among the 
finite
subgroups of $SU(3)$, there are two sets of subgroups called
type~C and type~D in~\cite{miller} that up to now defy a simple
general characterization. Groups of type~C and~D are best defined in terms of
generators~\cite{miller,ludl}:
\begin{subequations}
\begin{eqnarray}
\mbox{type C:} &&
E =
\left(
\begin{array}{ccc}
0 & 1 & 0\\
0 & 0 & 1\\
1 & 0 & 0\\
\end{array}
\right),
\label{E}
\;\;
F(\mu,\alpha,\beta) = \left(
\begin{array}{ccc}
\eta^\alpha & 0 & 0 \\ 0 & \eta^\beta & 0 \\ 0 & 0 & \eta^{-\alpha-\beta}
\end{array} \right)
\; \mbox{with} \; \eta = e^{2\pi i/\mu}, \hphantom{xx}
\\[2mm]
\mbox{type D:} &&
E, \;\; F(\mu,\alpha,\beta), \;\;
R(\nu,\rho,\sigma) =
\left( 
\begin{array}{ccc}
\delta^\rho & 0 & 0 \\ 0 & 0 & \delta^\sigma \\ 0 & -\delta^{-\rho-\sigma} & 0 
\end{array} \right)
\;\; \mbox{with} \; \delta = e^{2\pi i/\nu}.
\label{Dgenerators}
\end{eqnarray}
\end{subequations}
The numbers $\alpha$, $\beta$, $\mu$ ($0 \leq \alpha,\beta < \mu$) and 
$\rho$, $\sigma$, $\nu$ ($0 \leq \rho,\sigma < \nu$) are integers.

This rather opaque characterization of groups of type~C and~D becomes
elucidated by a theorem about Abelian subgroups $\mathcal{A}$ of
$SU(3)$ proven in~\cite{comments}, 
which states that for every $\mathcal{A}$ one can
find numbers $m$ and $n$, with $n$ being a divisor of $m$, such that
$\mathcal{A} \cong \zz_m \times \zz_n$. Since $E$ generates a $\zz_3$,
a group of type~C has the structure of a semidirect product
$\mathcal{A} \rtimes \zz_3$. As for type~D, one can find an
appropriate basis with two non-diagonal generators, given by the 
matrices $E$ and 
\begin{equation}\label{B}
B
:=
\left(
\begin{array}{rrr}
-1 &  0 &  0\\
 0 &  0 & -1\\
 0 & -1 &  0\\
\end{array}
\right).
\end{equation}
Since $E^3 = B^2 = (EB)^2 = \bone$, the group generated by $E$ and $B$
is $S_3$. Therefore, a group of type~D may be written as 
$\mathcal{A} \rtimes S_3$~\cite{review}.
In summary, the structures of groups of type~C and~D are given by
\begin{subequations}\label{CD}
\begin{eqnarray}
\mbox{type C:} && 
\gr \cong \left( \zz_m \times \zz_n \right) \rtimes \zz_3,
\label{C1}
\\
\mbox{type D:} && 
\gr \cong \left( \zz_m \times \zz_n \right) \rtimes S_3,
\label{D1}
\end{eqnarray}
\end{subequations}
respectively. In these formulas it is understood that $n$ is a divisor of
$m$. 

Equation~(\ref{CD}) characterizes the general structure, but neither
says for which $m$ and $n$ such groups exist nor gives a clue about the
generators of the normal subgroup $\zz_m \times \zz_n$.
In this paper we will show,
by an explicit construction of their generators,
that groups of type~C and~D are characterized by
three integers, namely $m$, $n$ and a third one which we denote by
$k$. We will derive a necessary condition which relates these three
numbers and which gives some information on the numbers $m$ and
$n$. With regard to groups of type~D, we will see that 
there is an additional condition on the three numbers.

Special cases of equation~(\ref{CD}) are 
well-known 
in the literature. 
For $m=n$ 
the corresponding group series are 
$\Delta(3n^2)$ 
(type~C) and   
$\Delta(6n^2)$ 
(type~D)~\cite{fairbairn,bovier1,luhn3n2,luhn6n2}.
In the case of $n=1$ and type~C, groups of the series $T_m$ exist only for
special $m$~\cite{bovier1,fairbairn1}. As we will see later, the $T_m$
have no counterpart of type~D.

The paper is organized as follows.
In section~\ref{groupsCD} we investigate the general structure of the
groups of type~C and~D, including questions concerning isomorphisms 
and factorization properties. 
The material in this section enables us to write down
all possible groups of type~C and~D.
In section~\ref{special} we present a complete list of the infinite series, in
which these groups are arranged.
We also mention some special cases 
as an illustration of the general results.
We summarize and conclude in
section~\ref{concl}. In the appendix we relate the notation 
in the present paper with that previously
introduced in~\cite{ludl,comments}
for uniquely characterizing groups of type~C and~D. 

\section{The groups of type C and D}
\label{groupsCD}
As demonstrated in~\cite{comments},
the essential ingredient for unveiling the structure of groups of
type~C and~D
is found
in the structure of Abelian
subgroups of $SU(3)$. Since both types of the groups under discussion
contain the matrix $E$ of equation~(\ref{E}), 
we will first prove a theorem concerning 
Abelian subgroups of $SU(3)$ which are invariant under
a similarity transformation with $E$.
In this proof, we will use the following lemma.
\begin{lemma}\label{divisorlemma}
Let $\mathcal{A}$ be an Abelian group 
and let $m$ be the highest order
among the orders of the elements of $\mathcal{A}$. Then the order $n$ of any
element of $\mathcal{A}$ is a divisor of $m$.
\end{lemma}
For the proof of this lemma we refer the reader to~\cite{comments}.
\begin{theorem}\label{abelian}
Let $\mathcal{A}$ be a finite subgroup of 
$SU(3)$ which consists of diagonal matrices and is invariant under the action
of the matrix $E$, \textit{i.e.}\ $E\mathcal{A}E^{-1}=\mathcal{A}$.
Then one can find positive integers $m$ and $n$, where $n$ is a divisor of
$m$, such that
\begin{equation}
\mathcal{A} \cong \zz_m \times \zz_n.
\end{equation}
Furthermore, 
the cyclic groups $\zz_m$ and $\zz_n$ are generated by matrices 
$F$ and $G$ of $\mathcal{A}$, respectively, which
have the form
\begin{equation}\label{FG}
F = \left( \begin{array}{ccc}
\epsilon & 0 & 0 \\ 0 & \epsilon^k & 0 \\ 0 & 0 & \epsilon^{-k-1} 
\end{array} \right),
\quad
G = \left( \begin{array}{ccc}
1 & 0 & 0 \\ 0 & \epsilon^{-r} & 0 \\ 0 & 0 & \epsilon^r
\end{array} \right)
\end{equation}
with 
$\epsilon = \exp (2\pi i/m)$, 
$k \in \{ 0,1,\ldots, r-1 \}$ and 
$r = m/n$.
\end{theorem}
\textbf{Proof:} 
Let $m$ be the
highest order of all matrices in $\mathcal{A}$.
Then there exists at least one matrix
$F_0\in\mathcal{A}$
of order $m$.
Due to $\mathrm{det}\,F_0=1$, this matrix has the form
\begin{equation}
F_0 = \mathrm{diag}\left(
\epsilon_1,\, \epsilon_2,\, (\epsilon_1 \epsilon_2)^{-1}\right),
\end{equation}
where $\epsilon_1$ and $\epsilon_2$ are roots of unity.
We denote the orders of $\epsilon_1$ and $\epsilon_2$ by $s_1$ and
$s_2$, respectively. If $t$ is the greatest common divisor of $s_1$
and $s_2$, then $s_1 = s_1't$, $s_2 = s_2't$ where $s'_1$ and $s'_2$
are coprime. 
Therefore, $\mathrm{ord}(F_0) = m = s_1's_2't$, and  
$\epsilon_1$ and $\epsilon_2$ 
generate a cyclic group of order $m$.
Hence, there are integers $a_1$ and $a_2$ 
($0 \leq a_i < s_i$ for $i=1,2$)
such that
\begin{equation}
\epsilon_1^{a_1} \epsilon_2^{a_2} = \exp(2\pi i/m)\equiv \epsilon,
\end{equation}
which implies that
\begin{equation}
F_1 \equiv  F_0^{a_1} (E F_0 E^{-1})^{a_2} =
\left( \begin{array}{ccc}
\epsilon & 0 & 0 \\ 0 & \epsilon^a & 0 \\ 0 & 0 & \epsilon^{-a-1} 
\end{array} \right)\in \mathcal{A}
\end{equation}
for some $a\in\{0,\,\ldots,\,m-1\}$.
Next we multiply all matrices in $\mathcal{A}$ with an
appropriate power of $F_1$ such 
that their first entries
become~1.
In this way
we obtain the subgroup of $\mathcal{A}$ of diagonal matrices
having 1 in the first entry. 
Obviously, this subgroup must be cyclic
and we denote its order by $n$.
According to lemma~\ref{divisorlemma},
$n$ is a divisor of $m$ and $\mathcal{A}$ has a generator $G$ as
stated in the theorem. 
Finally, we reconsider $F_1$. If $a \leq r-1$, then $k = a$ 
and $F = F_1$. If $a \geq r$, we multiply $F_1$ by an appropriate
power $b$ of $G$ such that 
$0 \leq a - br \leq r-1$. 
In this case, we define
$k = a - br$ and $F = F_1 G^b$.
This $F$ has the form as stated in the theorem. 
$\Box$

Theorem~\ref{abelian} will now enable us to prove the 
main result of the paper.
\begin{theorem}\label{main}
For every group $\gr$ of type~C, there exist positive integers $m$,
$n$ and $k$, where $n$ is a divisor of $m$ and $k$ is related to 
$r = m/n$ by
\begin{equation}\label{1}
1 + k + k^2 = 0 \mod r
\end{equation}
such that in a certain basis of\, $\mathbbm{C}^3$ its
generators are given by $E$, $F$ and $G$.
If the group is of type~D, one has to add
the matrix $B$ to the set of generators and
$k$ has to fulfill the further condition
\begin{equation}\label{2}
1 + 2k = 0 \mod r.
\end{equation}
\end{theorem}
The matrix 
$E$ is defined in equation~(\ref{E}),
$F$ and $G$ are the generators occurring in theorem~\ref{main}, and 
$B$ is given by equation~(\ref{B}). \\[1mm]
\textbf{Proof:} 
Let us begin with groups $\gr$ of type~C. 
Since $E \in \gr$, also $E M E^{-1} \in \gr$ for any matrix 
$M \in \gr$. This allows simultaneous cyclic permutations
in the normal subgroup $\mathcal{A} \subset \gr$ 
consisting of diagonal matrices.  
Therefore, theorem~\ref{abelian} is applicable and
there are elements $F$ and $G$ of the form of equation~(\ref{FG})
which generate $\mathcal{A}$.
Since
\begin{equation}
E F E^{-1} = F^k\, 
\mbox{diag} \left( 1, \epsilon^{-(1+k+k^2)}, \epsilon^{1+k+k^2}
\right) \in \mathcal{A}, 
\end{equation}
it follows that $\mathrm{diag}( 1, \epsilon^{-(1+k+k^2)}, \epsilon^{1+k+k^2})$
must be a power of $G$. Hence $1+k+k^2$ is a multiple of $r$.
In other words, 
equation~(\ref{1}) is a necessary 
condition that $\gr$ is an $SU(3)$-subgroup of type~C.
One can easily check that 
$E G E^{-1} \in \mathcal{A}$ is automatically fulfilled, thus 
equation~(\ref{1}) is also sufficient.

Switching now to subgroups of type~D, we find 
\begin{equation}
B F B^{-1} = F \,
\mbox{diag} \left( 1, \epsilon^{-(1+2k)}, \epsilon^{1+2k} \right),
\end{equation}
whence it follows that for such groups there is a second 
consistency condition given by equation~(\ref{2}). Since
$BGB^{-1} = G^{-1}$, no further condition ensues.
$\Box$

It is useful to summarize the action of $E$ and $B$ on the generators
$F$ and $G$ of the Abelian subgroup. To this end 
we reformulate equations~(\ref{1}) and~(\ref{2}) as 
\begin{equation}
\label{ell}
1 + k +k^2 = \ell r, \quad 1 + 2k = \ell' r,
\end{equation}
respectively, with positive integers $\ell$ and $\ell'$.
With these numbers the sought-after relations are given by
\begin{equation}\label{CDrelations}
E F E^{-1} = F^k G^\ell, \quad
E G E^{-1} = F^{-r} G^{-(k+1)}, \quad
BFB^{-1} = F G^{\ell'}, \quad
BGB^{-1} = G^{-1}.
\end{equation}
These equations, together with $F^m=G^n=\mathbbm{1}$, $FG=GF$
and the $S_3$-relations $E^3=B^2=(EB)^2=\mathbbm{1}$ comprise
a presentation of the groups of type~C and~D.

As we have seen so far, groups of type~C and~D are uniquely defined by
the triple $(m,n,k)$. Therefore, we can uniquely denote these groups
by the symbols $C_{m,n}^{(k)}$ and $D_{m,n}^{(k)}$, respectively. 
In theorem~\ref{main}, we have given a prescription which allows to
decide for which triples $(m,n,k)$ a group $C_{m,n}^{(k)}$ or a group 
$D_{m,n}^{(k)}$ exists---see equations~(\ref{1}) and~(\ref{2}). 
Clearly, if for such a triple a group of type~D
exists, then there is a corresponding group of type~C. 
Evidently, due to equation~(\ref{2}), the converse is not true in general.

One may ask the question for which $r$ it is possible to fulfill
equation~(\ref{1}). The following lemma provides the answer.
\begin{lemma}\label{pfd}
The expression $\phi(k) := 1 + k + k^2$ has 
the prime factor decomposition $3^i q$, where $i=1$ for 
$k = 1 \mod 3$, $i=0$ for $k = 0 \mod 3$ and $k = 2 \mod 3$, 
and 
$q = \prod_j (6z_j+1)$
is a product 
of primes of the form $6z_j+1$ with positive integers $z_j$.
Vice versa, the equation $\phi(k) = 0 \mod{r}$ has a solution whenever the
prime factor decomposition of $r$ is of the form above.
\end{lemma}
\textbf{Proof:}
We observe that $\phi(k)$ is identical with the third cyclotomic
polynomial $\Phi_3(k)$. The general theory of the prime factor decomposition
of cyclotomic polynomials $\Phi_n(k)$ says that any prime factor $p_j$ of
$\Phi_n(k)$ is either a divisor of $n$ or of the form 
$p_j = n z'_j + 1$~\cite{gallot,ribenboim}. 
Thus, in the case of $n=3$ we have $p_j=3$ or $p_j = 3 z'_j + 1$. However,
since $p_j$ is odd, $z'_j$ must be even and we arrive at $p_j = 6 z_j + 1$ with 
$z'_j = 2z_j$.
Finally, by direct computation it can be easily seen that the
factor 3 occurs in $\phi(k)$ iff $k = 1 \mod 3$, in which case it occurs
exactly once.
This proves the first part of the lemma.

Now we consider the equation $\phi(k) = 0 \mod{r}$.
If $r=q$ with $q$ as stated in the
lemma, it was proven in~\cite{bovier1}
that a solution exists.
It remains
to show that also for $r = 3q$ a solution exists. We know from the first part
of the lemma that this can only be the case if $k = 1 \mod 3$, \textit{i.e.} 
$k=3s+1$.
With this substitution we obtain  $\phi(k) =
3(1+3s+3s^2)$. Therefore, $s$ fulfills the equation 
$\phi_1(s) := 1+3s+3s^2 = 0 \mod{q}$.
Since 12 and $q$ are coprime, we can
multiply the left-hand side of 
this equation with 12 and obtain the equivalent equation
$(6s + 3)^2 + 3 = 0 \mod{q}$. The theory of congruences says that from a
solution of $t^2 + 3 = 0 \mod{q}$ we recover a solution of 
$\phi_1(s) = 0 \mod{q}$
via $6s + 3 = t \mod{q}$. So it remains to show 
that $t^2 + 3 = 0 \mod{q}$ has a solution. To this end we reconsider 
the general equation $\phi(l) = 0 \mod{q}$, 
of which it is indeed known that it has a
solution. Since 4 and $q$ are coprime, we can multiply the left-hand side of
this equation by 4 and obtain the equivalent equation
$(2l+1)^2 + 3 = 0 \mod{q}$. In other words, both 
$\phi_1(s) = 0 \mod{q}$ and $\phi(l) = 0 \mod{q}$ lead to the same 
equivalent simple quadratic congruence $t^2 + 3 = 0 \mod{q}$. Therefore,
this quadratic congruence has a solution, 
whence the same follows for $\phi_1(s) = 0 \mod{q}$.
$\Box$

Next we address the question, whether there are groups of type~C
or~D which factorize into direct products. Indeed such groups exist, but
this factorization is very limited according to the following theorem.
\begin{theorem}\label{factorization}
A group $C_{m,n}^{(k)}$ or $D_{m,n}^{(k)}$ factorizes if and only if 
$m = 3m'$ and neither $m'$ nor $n$ are divisible by 3. In this case,
$k=1 \mod{3}$ and 
the factorization is given by
\begin{equation}
C_{3m',n}^{(k)} \cong \zz_3 \times C_{m',n}^{(k)} 
\quad \mbox{or} \quad
D_{3m',n}^{(k)} \cong \zz_3 \times D_{m',n}^{(k)}.
\end{equation}
\end{theorem}
\textbf{Proof:} 
We prove the theorem for groups of type~C. Since any group of type~D
emerges from a group of type~C by adding the generator~$B$, the proof
holds for groups of type~D as well.

Suppose $\mathcal{G}$ factorizes as 
$\mathcal{G}\cong \mathcal{G}_1 \times \mathcal{G}_2$ 
with non-trivial $\mathcal{G}_1$ and $\mathcal{G}_2$. 
Since $\mathcal{G}$ contains non-diagonal matrices, 
not all elements of $\mathcal{G}_1$ and $\mathcal{G}_2$ are diagonal. 
Without loss of generality we assume that $\mathcal{G}_2$ 
contains an element of the form $ED$, where 
$D\in \mathcal{G}$ is a diagonal matrix. 
Due to the 
direct-product structure 
all elements of $\mathcal{G}_1$ commute with $ED$. 
Any diagonal matrix commuting with $ED$ must be proportional to the
unit matrix. Therefore, 
if $\mathcal{G}_1$ contains only diagonal matrices, we find 
$\mathcal{G}_1=\{\bone,\, \omega \bone,\, \omega^2\bone\}\cong \zz_3$
and we are finished. 
In the following we will discuss the case
that both $\mathcal{G}_1$ and $\mathcal{G}_2$ contain non-diagonal elements.

However, if $\mathcal{G}_1$ contains non-diagonal elements, at least one of
them has the form\footnote{If we assume that $\mathcal{G}_1$ contains
  an element of 
  the form $E^2 \tilde D$ with a diagonal matrix $\tilde D$, then
  $(E^2 \tilde D)^2$ is of the stated form.} 
$ED'$ with a diagonal matrix $D'\in \mathcal{G}$. 
It is easy to convince oneself 
that $[ED,ED']=0$ implies $D'= \lambda D$, $\lambda\in\mathbbm{C}$ for any
diagonal matrices $D$ and $D'$, a fact 
which we will frequently use in the following. Consequently, $ED' = \lambda
ED$ and thus  
\begin{equation}\label{lambda}
\lambda\bone = (ED')(ED)^{-1} \in\mathcal{G} \quad\text{and}\quad
\lambda^2\bone = (ED')^2(ED)^{-2}\in\mathcal{G}.
\end{equation}
We note that $\lambda\neq 1$, because else $ED=ED'$ which is a contradiction to
$\mathcal{G}_1\cap\mathcal{G}_2=\{\bone\}$. 
Therefore, $\lambda=\omega$ or $\lambda=\omega^2$ due to 
$\det (\lambda \bone) = 1$. 
Suppose now that there is another matrix $ED''$ ($D''$  diagonal) in
$\mathcal{G}_1$. 
Using the same argument as above, we would find $D''=\tilde\lambda D$
$\Rightarrow\tilde\lambda\bone = (ED'')(ED)^{-1}\in\mathcal{G}$
with $\tilde\lambda=\omega$ or $\tilde\lambda=\omega^2$, \textit{i.e.}\ we
find a decomposition 
$\tilde\lambda\bone = g_1 g_2$ with $g_1\in\mathcal{G}_1,\,g_2\in\mathcal{G}_2$.
Since $\mathcal{G}\cong\mathcal{G}_1\times\mathcal{G}_2$, this decomposition
must 
be unique, but this leads to a contradiction since equation~(\ref{lambda})
already 
comprises a different decomposition of this type. Therefore, $ED''=ED'$.
The same argument is equally applicable to $\mathcal{G}_2$, \textit{i.e.}\ we
find that there also cannot be a matrix $ED'''$ in $\mathcal{G}_2$.
Similarly one can deduce that the matrices $(ED')^2$ and $(ED)^2$ are the only
matrices containing $E^2$ in $\mathcal{G}_1$ and $\mathcal{G}_2$, respectively.
Since any diagonal matrix which commutes with $ED$ or $ED'$ must be
proportional to 
$\bone$, we find that the diagonal matrices in $\mathcal{G}_1$ and
$\mathcal{G}_2$ 
can only be $\omega^i\bone$ $(i=0,1,2)$. However, $\omega\mathbbm{1}$ and
$\omega^2\mathbbm{1}$ cannot be in $\mathcal{G}_1$ and $\mathcal{G}_2$,
because else 
$ED'\in\mathcal{G}_2$ which is a contradiction. Hence the groups
$\mathcal{G}_1$ and $\mathcal{G}_2$ are given 
by\footnote{Note that $(ED)^3 = \bone$ for any diagonal matrix $D \in SU(3)$.}
\begin{equation}\label{12*}
\mathcal{G}_1 = \{\bone,\, \lambda ED,\, \lambda^2 (ED)^2\} \quad\text{and}\quad
\mathcal{G}_2 = \{\bone,\, ED,\, (ED)^2\}
\end{equation}
with $\lambda=\omega$ or $\lambda=\omega^2$. 
We finally exploit the fact that $E\in\mathcal{G}$.
This implies that also $D \in \mathcal{G}$. However, by
multiplication of the matrices occurring in equation~(\ref{12*}) we
can only produce diagonal matrices of the form $\omega^i\bone$. Hence
we find $D=\omega^i\bone$ ($i=0,1$ or $2$). So we end up with the only
possibility 
\begin{equation}
\mathcal{G}=\{ \omega^i E^j | i,j=0,1,2\}\cong \zz_3\times \zz_3\cong
\{\bone,\, \omega \bone,\, \omega^2\bone\} \times \{\bone,\, E,\, E^2\},
\end{equation}
\textit{i.e.}\ also in this case we can choose
$\mathcal{G}_1=\{\bone,\, \omega \bone,\, \omega^2\bone\}\cong\zz_3$.
Consequently, if $\gr$ factorizes, then $\omega \bone \in \gr$. 
This is only possible if 3 is a divisor of $m$.

Thus, for the remainder of the proof, we depart from
\begin{equation}
\gr \cong \zz_3 \times \gr'
\end{equation}
and investigate $\gr'$. We know that $\omega \bone \not\in \gr'$. Let us now
assume that $\gr'$ contains a diagonal matrix
$A$ whose order is $x = 3y$. In this case, 
\begin{equation}
A^y = \mbox{diag} \left( \omega^\alpha, \omega^\beta, \omega^\gamma \right)
\equiv A' \in \gr'
\end{equation}
with $\alpha + \beta + \gamma = 0 \mod{3}$. 
We know already that $\alpha = \beta = \gamma \neq 0$ is not possible. All
other possibilities are of the form $\alpha = 0$, $\beta = 1$, $\gamma = 2$ or
permutations thereof. Let us consider for instance
$A' = \mbox{diag} \left( 1, \omega, \omega^2 \right) \in \gr'$.
However, this leads to
\begin{equation}
\left( A' \right)^2 E A' E^{-1} = \omega \bone \in \gr',
\end{equation}
which is a contradiction.\footnote{If $E\not\in\gr'$, this
argument still holds with any matrix of the form $ED$ with a diagonal
matrix $D$. 
Since $E \in \gr \cong \zz_3 \times \gr'$, at least one such element
is contained in $\gr'$.} 
We conclude that $\gr'$
contains no diagonal element whose
order is divisible by 3.

Any diagonal matrix $D \in \gr$ must have a unique decomposition as 
$D = \omega^x D_1$ with $D_1 \in \gr'$. We know that 3 is not a divisor of the
order of $D_1$. Therefore, the order of
any diagonal element in $\gr$ contains the
prime factor 3 at most once.
Since $F \in \gr$, we conclude that $m' =m/3$
does not have 3 in its prime factor decomposition.
Since $G_{11}=1$, its decomposition is $G=\omega^0 G$, 
\textit{i.e.}\ $G\in \gr'$.
Hence, its order $n$ is not divisible by three.
With the same reasoning we also conclude that 
$F^3 \in \gr'$.
In summary we know that neither $m'$ nor $n$ are divisible by 3. 
Therefore, we find $r = 3r'$, with $r' = m'/n$ also not divisible by 3.
Then, according to lemma~\ref{pfd} we obtain 
\begin{equation}\label{13}
k=1 \mod{3}.
\end{equation}
This finishes the first part of the proof: Assuming that the group
factorizes we obtain the results as stated in the theorem.

Now we show the converse: If $m=3m'$ and neither $m'$ nor $n$
are divisible by three, the group factorizes. 
In order to prove this, we define a subgroup $\gr'$ of $\gr$ such
that every element of $\gr$
can uniquely be decomposed as $\omega^a X$ with $X\in \gr'$.
We define $\gr'$ as being generated by $F^3$, $G$ and $E$; in the case
of type~D we have to add the generator $B$.
Clearly the decomposition of $G$, $E$ and $B$ is unique with
$\omega^0$. 
To complete the proof we need to demonstrate that $F$
can be uniquely decomposed as 
$F = \omega^a F^{3b}$.
For this purpose, we consider 
the~11 and 22-elements in this matrix relation, which give
\begin{equation} 
\epsilon = \omega^a \epsilon^{3b}
\quad \mbox{and} \quad
\epsilon^k = \omega^a \epsilon^{3bk},
\end{equation}
respectively.
It is always possible to find integers $a$ and $b$ such that the first
relation holds, because 3 and $m'= m/3$ are coprime. 
Then, taking the $k$-th power of the first relation and comparing with the
second one, we find the consistency condition $\omega^{ka} = \omega^a$.
However, due to equation~(\ref{13}), this is fulfilled for any $a$.
Finally, the equation referring to the 33-element in $F = \omega^a F^{3b}$
is automatically true because we are dealing with $SU(3)$ matrices.

Therefore, 
$F^3$, $G$ and $E$ in the case of type~C, 
or both $E$ and $B$ in the case of type~D,
generate $\gr'$, with $k$ fulfilling
equation~(\ref{1}), or both equations~(\ref{1}) and 
equation~(\ref{2}). 
It remains to show that $\gr'$ can be written in
the form as stated in the theorem. Firstly, we note that 
the relations
\begin{equation}
1 + k + k^2 = 0 \mod{r'}
\quad \mbox{and} \quad
1 + 2 k = 0 \mod{r'}
\quad \mbox{with} \quad r' = r/3
\end{equation}
hold because they are valid for $r = 3r'$. Secondly, because of
$\epsilon^r = \left( \epsilon^3 \right)^{r'}$, the generator $G$ can
be conceived as being based on the root of unity 
$\epsilon^3 = \exp \left( 2\pi i/m' \right)$.
The same is
evidently true also for $F^3$. 
Thus $\gr'$ is given by $C_{m',n}^{(k)}$ or $\gr' = D_{m',n}^{(k)}$
with $r' = m'/n$.
It could happen, however, that $k \geq r'$. In this case, 
one has to multiply $F^3$ with a suitable power of $G$
in order to switch from $k$ to a $k' < r'$, just as we 
proceeded at the end of the proof of theorem~\ref{abelian}. This object is
then the new generator $F'$ in $\gr'$.
$\Box$

Since for groups $D_{m,n}^{(k)}$ the number $k$ has to be a
simultaneous solution of both equations~(\ref{1}) and~(\ref{2}),
the possible triples $(m,n,k)$ are 
considerably rarer than in the case of groups of type~C.
This is corroborated by the following theorem.
\begin{theorem}\label{rD}
A group $D_{m,n}^{(k)}$ exists only for 
\begin{enumerate}
\renewcommand{\labelenumi}{\alph{enumi})}
\item
$k=0$ with $m=n$, $r=1$,
\item
$k=1$ with $m = 3n$, $r=3$.
\end{enumerate}
\end{theorem}
\textbf{Proof:}
Multiplying the left-hand sides of equations~(\ref{1}) and~(\ref{2}) with 4 and
$1+2k$, respectively, we obtain
\begin{equation}
4(1+k+k^2) = 0 \mod{r} 
\quad \mbox{and} \quad
(1+2k)^2 = 0 \mod{r}.
\end{equation}
Subtracting these relations from each other, we find 
$3 = 0 \mod{r}$. Therefore, $r=1$ or $r=3$ which leads to $k=0$ or $k=1$,
respectively. $\Box$
\\[2mm]
The theorem says that
all groups of type~D
belong to two classes: 
$D_{n,n}^{(0)}$ and $D_{3n,n}^{(1)}$ for arbitrary $n \geq 1$.
However, applying theorem~\ref{factorization}, we establish that
$D_{3n,n}^{(1)} \cong \zz_3 \times D_{n,n}^{(0)}$, whenever $n$ is not
divisible by 3, in which case the second factor is a group of the first
class.\footnote{Note that $D_{n,n}^{(0)} = D_{n,n}^{(1)}$ with the
argument at the end of the proof of theorem~\ref{factorization}.} This leaves $D_{9n',3n'}^{(1)}$ 
as the sole genuine series of groups of the second class.  
In summary, leaving out the case in which the group factorizes,
all groups of type~D are described by just two infinite
series, which are explicitly given
by
\begin{equation}\label{D}
D_{m,m}^{(0)} \;\; (m=1,2,3,\ldots) 
\quad \mbox{and} \quad 
D_{9n',3n'}^{(1)} \;\; (n' = 1,2,3,\ldots).
\end{equation}

Another interesting point is whether some of the groups 
$C_{m,n}^{(k)}$ are isomorphic. Clearly,
this can only happen for identical pairs $(m,n)$. Indeed, for one $r = m/n$
there can be several solutions $k_j$ of equation~(\ref{1}). 
The following theorem provides a 
necessary and sufficient 
condition for such an isomorphism.
\begin{theorem}\label{isomorphic}
Two groups $C_{m,n}^{(k_1)}$ and $C_{m,n}^{(k_2)}$ with $k_1\neq k_2$ are
isomorphic if and only if
\begin{equation}\label{iso-condition}
1+k_1+k_2=r.
\end{equation}
\end{theorem}
Before we present the proof, some comments are at order. 
\begin{enumerate}
\renewcommand{\labelenumi}{\roman{enumi}.}
 \item It is easy to see that, if $k_1$ is a solution of $1+k+k^2 = 0 \mod{r}$,
 so is $k_2 = r-k_1-1$.
 \item If $k_1$ and $k_2$ are equal, then $r=1+2k_1$, which implies $r=1$ or
   $r=3$---see the proof of theorem~\ref{rD}. 
 In these cases one finds the unique solutions $k_1=k_2=0$ and 
 $k_1=k_2=1$, respectively. 
 \item Consequently, for all $r>3$, the solutions of $1+k+k^2 = 0 \mod{r}$
   come in \textit{pairs} $(k_1,k_2)$. 
 Theorem~\ref{isomorphic}, however,
 tells us that the groups $C_{m,n}^{(k_1)}$ and $C_{m,n}^{(k_2)}$
 corresponding to such a pair are isomorphic, 
 though, as sets of matrices, they are not identical.
\end{enumerate}
\textbf{Proof:}
The smallest possible $r$ for which there are two solutions of $1+k+k^2 = 0
\mod{r}$ is $r=7$.  
Thus, for this proof, we can assume 
$m\geq 7$.

In the following we will always consider the two groups
$C_1=C_{m,n}^{(k_1)}$ and $C_2=C_{m,n}^{(k_2)}$ $(k_1\neq k_2)$, 
which have the same
generators $E$ and $G$, but different generators
$F_1$ with $k_1$ and $F_2$ with $k_2$, respectively.
Moreover, in the first relation in equation~(\ref{ell}), 
we now have $\ell_1$ and $\ell_2$ corresponding to $k_1$ and $k_2$,
respectively. 

Suppose that there exists an isomorphism $f: C_1 \rightarrow C_2$, then
$\mathrm{ord}(F_1)=m$ implies 
$\mathrm{ord}(f(F_1))=m\geq 7$. Thus, $f(F_1)$ cannot be a matrix of the
form $EA$ or $E^2A$, with $A$ being a diagonal matrix in $C_2$, 
because all such matrices are of order three. Similarly, $f(G)$ must also
be diagonal, because else $f(F_1)f(G)=f(G)f(F_1)$ 
would imply $f(F_1)\propto \bone$, which is not possible for $m>3$.
Therefore, an
isomorphism necessarily maps $F_1$ and $G$ 
into
diagonal matrices, 
\textit{i.e.}\ $f$ is of the form
\begin{equation}\label{generalform-iso}
f:
\begin{array}{ccl}
F_1 & \mapsto & F_2^\alpha G^\beta,\\
G   & \mapsto & F_2^\gamma G^\delta,\\
E   & \mapsto & EA \enspace\text{or}\enspace E^2A \quad (\text{$A$
  diagonal}). 
\end{array}
\end{equation}
The condition $\mathrm{ord}(f(F_1))=m$ implies that $\alpha$ and $r$ are
coprime. In 
the same manner $\mathrm{ord}(f(G))=n$ enforces $\gamma$ to be a multiple
of $r$, 
\textit{i.e.}\ $\gamma=\gamma' r$. 
Applying $f$ to the group
relation $EF_1 E^{-1} = F_1^{k_1} G^{\ell_1}$ we obtain the consistency
condition 
\begin{equation}
f(E) f(F_1) f(E)^{-1} = f(F_1)^{k_1} f(G)^{\ell_1}.
\end{equation}
We will first study the case $f: E\mapsto E^2 A$. Using
equation~(\ref{generalform-iso}), we 
obtain
\begin{equation}
E^2 F_2^\alpha G^\beta E^{-2} = F_2^{\alpha k_1 + \gamma' r\ell_1} 
G^{\beta k_1 +\delta \ell_1}. 
\end{equation}
The left-hand side of this equation can be simplified by means of the
relations~(\ref{CDrelations}) 
for the group $C_2$. 
In doing so 
one finds
\begin{equation}
F_2^{\alpha (k_2^2-k_1)} = F_2^{ra} G^b
\end{equation}
with 
$a = \alpha \ell_2 - \beta +\gamma' \ell_1$ and
$b = \alpha \ell_2 + \beta (k_1 - (1+k_2)^2 +r \ell_2 ) +\delta
\ell_1$.
Taking the 11-component of this matrix-equation we finally get
\begin{equation}
\epsilon^{\alpha (k_2^2 -k_1)} = \epsilon^{r a}.
\end{equation}
Since $\alpha$ and $r$ are coprime, it
follows that 
\begin{equation}
k_2^2-k_1 = 0 \mod{r}.
\end{equation}
Reformulating this by using equation~(\ref{1}), one finds
\begin{equation}
1 + k_1 + k_2 = 0 \mod{r}.
\end{equation}
Since $0\leq k_1 \leq r-1$ and $0\leq k_2 \leq r-1$, this implies
$1 + k_1 + k_2 = r$, \textit{i.e.}\ equation~(\ref{iso-condition}).
Repeating the same procedure for isomorphisms of the form
$f: E \mapsto EA$, one obtains the trivial condition $k_1=k_2$.

It remains to show that equation~(\ref{iso-condition}) 
is also sufficient for $C_1$ and $C_2$ to be isomorphic.
This can be done by explicitly 
specifying
an isomorphism.
Because of $1+k_1+k_2 = r$, we can write $F_1$ as
\begin{equation}
F_1 = \mbox{diag} \left( \epsilon, \epsilon^{-1-k_2+r},
\epsilon^{k_2-r} \right),
\end{equation}
resulting in the relation
\begin{equation}
B \left( F_1 G \right) B^{-1} = F_2
\quad \mbox{or} \quad
B F_1 B^{-1} = F_2  G,
\end{equation}
where $B$ is defined in equation~(\ref{B}). Since also the relations
$B E B^{-1} = E^2$, $B G B^{-1} = G^{-1}$ hold, 
the mapping 
\begin{equation}
M \in C_{m,n}^{(k_1)} \to BMB^{-1} \in C_{m,n}^{(k_2)}
\end{equation}
procures an isomorphism between the two groups. 
$\Box$
\\[2mm]
Since groups of type~C occur in isomorphic pairs, it is sufficient to consider
in each pair the group with the smaller $k$. It can happen, however, that
equation~(\ref{1}) has several pairs of solutions $(k_1,k_2)$, related by 
$1+k_1+k_2 = r$. 
The first $r$ where this happens is $r=91$, with two pairs 
of solutions 
$(k_1,k_2) = (9,81)$ and $(k_1,k_2) = (16,74)$.
From theorem~\ref{isomorphic} we know that
$C_{91n,n}^{(9)}$ and $C_{91n,n}^{(16)}$ are indeed non-isomorphic for all
$n=1,2,3,\ldots$. 
The next values of $r$ showing this 
feature are $r=133$ and $r=217$.

\section{Infinite series of groups of type~C and~D}
\label{special}
\begin{table}
\begin{center}
\begin{tabular}{rrcrr}
$r$ & $k$ & \hphantom{xxxx} & $r$ & $k$ \\\cline{1-2} \cline{4-5}
\rule{0mm}{5.5mm} 
1 & 0                   && $7 \times 7 = 49$ & 18 \\
3 & 1                   && $3 \times 19 = 57$ & 7 \\
7 & 2                   && 61 & 13 \\
13 & 3                  && 67 & 29 \\
19 & 7                  && 73 & 8 \\
$3\times 7 = 21$ & 4    && 79 & 23 \\
31 & 5                  && $7 \times 13 = 91$ & 9,\,16 \\
37 & 10                 && $3 \times 31 = 93$ & 25 \\
$3 \times 13 = 39$ & 16 && 97 & 35 \\
43 & 6                  &&&
\end{tabular}
\end{center}
\caption{List of all allowed $r< 100$ and the corresponding
  solution(s) $k$ of equation~(\ref{1}). Also shown is the prime
  factor decomposition of $r$.
\label{r-table}}
\end{table}
In order to obtain a comprehensive impression of 
the existing infinite series of 
groups of type~C and~D, it is at first useful to consider the values of $r$. 
In table~\ref{r-table} we have listed all allowed $r < 100$ in ascending
order.
Clearly, these values of $r$ are
in agreement with lemma~\ref{pfd}.
In table~\ref{r-table} also the 
solutions of equation~(\ref{1}) for a
given $r$ are shown. 
For details on these solutions, which are relevant for the table, we
refer the reader to the end of the previous section.

Infinite series of groups are obtained
if we fix $r$ and vary $n$, in which case $m$ varies as $m = rn$. 
In this way, for groups of type~C, there is an infinite series 
$C_{rn,n}^{(k)}$ ($n=1,2,3,\ldots$) for every pair $(r,k)$ where $k$
is related to $r$ via equation~(\ref{1}). 
However, in view of the factorization theorem~\ref{factorization},
it is meaningful 
to distinguish three cases:
\begin{enumerate}
\renewcommand{\labelenumi}{\alph{enumi})}
\item
3 does not divide $r$: In this case we simply obtain 
$C_{rn,n}^{(k)}$ ($n=1,2,3,\ldots$).
\item
3 divides $r$ but not $n$: Writing $r = 3r'$,
according to lemma~\ref{pfd}, $r'$ is not divisible by 3. 
The factorization theorem~\ref{factorization} applies and 
we arrive at the series $\zz_3 \times C_{r'n,n}^{(k)}$. Removing the trivial
factor $\zz_3$, we are lead back to case a).
\item
3 divides both $r$ and $n$: The factorization theorem is not applicable
and we obtain the series $C_{9r'n',3n'}^{(k)}$ ($n'=1,2,3,\ldots$), where
we have written $n = 3n'$. 
\end{enumerate}
Note that, according to
theorem~\ref{rD}, only $r=1,3$ have also series of type~D. 

For $r=1$, which implies $k=0$,
the two series of type~C and~D are the 
well-known 
dihedral-like groups~\cite{fairbairn} 
\begin{equation}\label{Delta}
C_{n,n}^{(0)} \equiv \Delta(3n^2) 
\cong \left( \zz_n \times \zz_n \right) \rtimes \zz_3
\quad \mbox{and} \quad 
D_{n,n}^{(0)} \equiv \Delta(6n^2)
\cong \left( \zz_n \times \zz_n \right) \rtimes S_3,
\end{equation}
respectively. 
Obviously, $n=1$ gives the trivial cases $\zz_3$ and $S_3$, respectively,
while for $n=2$ there are the
well-known isomorphisms
$\Delta(12) \cong A_4$ and $\Delta(24) \cong S_4$.
For the groups of equation~(\ref{Delta}),
the generators of the Abelian subgroup are related by 
$G = EFE^{-1}$~\cite{luhn3n2,luhn6n2}. 
Therefore, in this case one of these two generators is actually
superfluous. 

Along this line of reasoning,
one could speculate if one diagonal generator
is always sufficient for groups of type~C and~D. 
Actually, this is suggested by the parameterization of groups of
type~C in the form of equation~(\ref{E}), which has indeed only one diagonal
generator. So if this parameterization comprises \emph{all} groups of type~C
studied in 
the present paper, then it should always be
possible to construct a single 
diagonal generator as a product of powers of $F$ and $G$ which could replace
both $F$ and $G$. Since every group of type~D can be conceived as a semidirect
product of a group of type~C with the $\zz_2$ generated by~$B$ of
equation~(\ref{B}), one diagonal generator should also be sufficient in this
case. Actually, we have verified this numerically for both type~C and~D for
all $m\leq 3000$, but we could not prove this in general.

Next we discuss $r=3$, which implies $k=1$, with the series
\begin{equation}\label{CDseries}
C_{9n',3n'}^{(1)} \cong
\left( \zz_{9n'} \times \zz_{3n'} \right) \rtimes \zz_3
\quad \mbox{and} \quad 
D_{9n',3n'}^{(1)} \cong
\left( \zz_{9n'} \times \zz_{3n'} \right) \rtimes S_3
\end{equation}
for $n'=1,2,3,\ldots$, belonging to case c) above. 
The smallest group of this type C-series is~\cite{comments}
\begin{equation}
C_{9,3}^{(1)} \cong 
\left( \zz_9 \times \zz_3 \right) \rtimes \zz_3
\end{equation}
with order 81. 
The smallest members of this
type D-series, which
recently have received some attention as 
flavour groups~\cite{lindner,holthausen,hagedorn,levaillant}, are
\begin{equation}
D_{9,3}^{(1)} \cong \left( \zz_9 \times \zz_3 \right) \rtimes S_3
\quad \mbox{and} \quad
D_{18,6}^{(1)} \cong \left( \zz_{18} \times \zz_6 \right) \rtimes S_3
\end{equation}
with orders 162 and 648, respectively.

The remaining series are all of type~C:
\begin{eqnarray}
C_{rn,n}^{(k)}\;\; (n=1,2,3, \ldots),
&&
r \; \mbox{coprime to}\; 3,
\label{c2} \\
C_{9r'n',3n'}^{(k)}\;\; (n'=1,2,3, \ldots),
&&
r' \; \mbox{coprime to}\; 3. 
\label{c1} 
\end{eqnarray}

The first members of the series of equation~(\ref{c2}) with $n=1$, 
which were first discussed in~\cite{bovier1,fairbairn1}, 
are denoted by 
\begin{equation}
C_{m,1}^{(k)} \equiv T_m \cong \zz_m \rtimes \zz_3.
\end{equation}
Since $r=m$, the equation for $k$ is
\begin{equation}
1+k+k^2 = 0 \mod m.
\end{equation}
Using
table~\ref{r-table}, we can read off the possible values of
$r=m$.
Since $r$ must be coprime to 3, we have to leave out all values of $r$
which have a number 3 in their prime factor decomposition.
The values $m=r$
for the smallest groups $T_m$ 
are all primes of the form $6z+1$, in
agreement with lemma~\ref{pfd}. 
The smallest group, where $m$ is a product of
such primes, is $T_{49}$. 
As discussed at the end of section~\ref{groupsCD}, 
for $m=91=7 \times 13$ the
label $T_{91}$ is ambiguous~\cite{review}, because it denotes both
non-isomorphic groups $C_{91,1}^{(9)}$ and $C_{91,1}^{(16)}$.
These two groups are the first members of the two series
$C_{91n,n}^{(9)}$ and $C_{91n,n}^{(16)}$, respectively, of non-isomorphic
groups with the same $m$ and $n$ but different values of $k$.

\section{Conclusions}
\label{concl}
In this paper we have studied the subgroups of $SU(3)$ that are called
type~C and~D in~\cite{miller}. 
We have proven five theorems in section~\ref{groupsCD} whence it follows 
that every group of type~C or~D is a member
of a series of groups which has infinitely many members.
Interestingly, the groups of type~D are arranged in only two such series whereof
one is the well-known
series $\Delta(6n^2)$.\footnote{In~\cite{zwicky} it was shown that 
every group of type~D is a subgroup
of an appropriate dihedral-like group. This is confirmed by
our results because for the second series the relation
$D_{9n',3n'}^{(1)} \subset \Delta(6 \times (9n')^2)$ holds.}
In contrast to this, the groups of type~C comprise
infinitely many series, two of which being the well
known series $\Delta(3n^2)$ and $T_m$. Our findings agree with the list of
groups up to order 512 in~\cite{subgroups-u3} which has been constructed using
GAP~\cite{GAP} and the library of small groups~\cite{SGL}.

Theorem~\ref{main} is the main result of this paper. It shows that groups of
type~C and~D can uniquely be characterized by three numbers $m$, $n$ and $k$.
Thus we have denoted the groups by $C_{m,n}^{(k)}$ and $D_{m,n}^{(k)}$,
with orders $3mn$ and $6mn$, respectively. The general structure of these
groups is given by equation~(\ref{CD}). 
The three numbers are related in the following way: 
$m = rn$ and $k$ is a solution of $1+k+k^2 = 0 \mod{r}$. The numbers $k$ and
$r$ fix the generators $F$ and $G$ according to theorem~\ref{main},
while 
the generators $E$ and $B$ of equations~(\ref{E}) and~(\ref{B}), respectively,
are independent of these numbers. In our scheme, groups of type~C have three
generators, $F$, $G$ and $E$, for type~D there is one more generator $B$.

We have found a systematic way
for constructing all infinite series of both types. 
While for type~D we could do this explicitly in view of only two such
series, for type~C one can choose the following procedure. Considering 
equation~(\ref{1}), one can write down those values of $r$, up to
arbitrarily high numbers, for which a solution $k$ exists; in this context
lemma~\ref{pfd} comes to aid. 
Then, for each $r$, an infinite series is obtained
by varying $n$. Details of this procedure depend on whether 3 is a
divisor of $r$ or not. 
Since there are infinitely many admissible $r$, for type~C infinitely
many series are obtained. The case of type~D is much more restricted,
allowing only $r=1$ and 3.
We have summarized our main result, 
the complete list of series of the groups of type~C and~D, 
in table~\ref{CDsummary}.

In conclusion, 
we have found a systematic way for computing the
admissible numbers $m$ and $n$ in equation~(\ref{CD}).
An essential ingredient
for achieving this was the explicit construction of the generators of 
the Abelian subgroup $\zz_m \times \zz_n$ of the groups
of type~C and~D.

\begin{table}
\renewcommand{\arraystretch}{1.2}
\begin{center}
\begin{tabular}{|l|l|}
\hline
series & comments \\
\hline
$C_{rn,n}^{(k)}$ & $m=rn$, $3 \nmid r$, $n=1,2,3,\ldots$ 
\rule{0mm}{5.5mm}\\
$\zz_3\times C_{r'n,n}^{(k)}$ & $m = 3r'n$, $r=3r'$, $3 \nmid r'$, $3 \nmid n$ 
$\Rightarrow$ $n=1,2,4,5,7,\ldots$ \\
$C_{9r'n',3n'}^{(k)}$ & $m = 9r'n'$, $n=3n'$, $r=3r'$, $3 \nmid r'$, 
$n'=1,2,3,\ldots$ \\[1mm] 
\hline
$D_{n,n}^{(0)} \equiv \Delta(6n^2)$ & $m=n$, $r=1$ 
\rule{0mm}{5.5mm}\\
$\zz_3 \times \Delta(6n^2)$ & $m=3n$, $r=3$, $3 \nmid n$ 
$\Rightarrow$ $n=1,2,4,5,7,\ldots$ \\
$D_{9n',3n'}^{(1)}$ & $m = 9n'$, $n=3n'$, $r=3$, $n'=1,2,3,\ldots$ \\[1mm]
\hline
\end{tabular}
\end{center}
\caption{Summary of the group series of type~C and~D.
The symbol $\nmid$ signifies
``does not divide.''
Concerning the groups of type~C in the first three lines, we note that, 
in the first line, $r$ can be any product of primes of the form
$6z+1$, whereas in the second and third lines this is true for $r'$.
The relation between $r$ and $k$ is given by equation~(\ref{1}).
There is an infinite series for every pair $(r,k)$.
Finally, we mention the special cases 
$C_{n,n}^{(0)} \equiv \Delta(3n^2)$ and $C_{m,1}^{(k)} \equiv T_m$ with $r=m$.
}
\label{CDsummary}
\end{table}

\vspace{5mm}

\noindent
\textbf{Acknowledgement:} This work is supported by the Austrian
Science Fund (FWF), Project No.\ P~24161-N16.

\newpage

\begin{appendix}

\section{Previously used notation for the groups of type C and D}

Since in this paper we have introduced a new systematic notation for the groups
of type~C and~D, we want to compare it with the notation put forward
in~\cite{ludl,subgroups-u3}. In this notation the 
groups of type~C are denoted by 
$C(\mu,\alpha,\beta)$, 
generated by the matrices of
equation~(\ref{E}). Similarly, groups of type~D are denoted by
$D(\mu,\alpha,\beta;\nu,\rho,\sigma)$, 
generated by the matrices of equation~(\ref{Dgenerators}).
In~\cite{comments} a prescription has been given to determine the
order and structure of any group of type~C when stated in the form 
$C(\mu,\alpha,\beta)$.
However, this procedure does not immediately tell us the value of
$k$ and we have not succeeded in finding a general formula which determines
$k$ as a function of $\mu$, $\alpha$ and $\beta$.
For all practical purposes, the best way to ``translate'' 
$C(\mu,\alpha,\beta)$
into the notation established in this paper is to explicitly perform the steps
described in the proof of theorem~\ref{abelian}. We shall concisely illustrate
this by means of an example.

Consider the group $C(\mu,\alpha,\beta)=C(28,4,22) \cong C_{m,n}^{(k)}$. 
We first have to find a diagonal matrix
of the highest possible order $m$. According to~\cite{comments} the
generator $F(\mu,\alpha,\beta)$ fulfills 
this requirement, thus in our example
\begin{equation}
m = \mathrm{ord}(F(28,4,22)) = 14.
\end{equation}
In the notation of the proof of theorem~\ref{abelian}, the matrix $F_0$ can
thus be chosen to be
\begin{equation}
F_0 = F(28,4,22) = \mathrm{diag}\left(\epsilon^2,\, \epsilon^{11},\,
\epsilon\right) 
\end{equation}
with $\epsilon=\exp(2\pi i/14)$.
In the next step we form $F_1 = F_0^{a_1} (E F_0 E^{-1})^{a_2}$
such that the first entry of $F_1$ is $\epsilon$, \textit{e.g.}
\begin{equation}
F_1 = F_0^0 (E F_0 E^{-1})^{9} = \mathrm{diag}\left(\epsilon,\, \epsilon^{9},\, \epsilon^{4}\right).
\end{equation}
Now we have to multiply all diagonal matrices in the group 
with a power of $F_1$ such 
that the 11-elements of the resulting matrices are all~1.
All diagonal elements of the group can be
expressed as products of
\begin{equation}
F(28,4,22) = \mathrm{diag}\left(\epsilon^2,\, \epsilon^{11},\, \epsilon\right) =
F_1^2\, \mathrm{diag}\left(1,\, \epsilon^7,\, \epsilon^{-7}\right)
\end{equation}
and
\begin{equation}
E F(28,4,22) E^{-1} = \mathrm{diag}\left(\epsilon^{11},\, \epsilon,\,
\epsilon^2 \right) = 
F_1^{11}\, \mathrm{diag}\left(1,\, 1,\, 1\right).
\end{equation}
Thus the group $\zz_n$ is generated by
\begin{equation}
G =  \mathrm{diag}\left(1,\, \epsilon^{-7},\, \epsilon^{7}\right) =
\mathrm{diag}\left(1,\, -1,\, -1\right). 
\end{equation}
Thus $n=2$ and $r=7$, and with equation~(\ref{1}) we obtain $k=2$. 
Finally, we have to form $F=F_1 G^b$ with a suitable power $b$
such that $F_{22}=\epsilon^k$; in our example we find
$F=F_1 G$.
Thus we arrive at the result 
\begin{equation}
C(28,4,22) = C_{14,2}^{(2)}.
\end{equation}

If one has a group of type~D given in the form 
$D(\mu,\alpha,\beta;\nu,\rho,\sigma)$, 
one first has to consider all diagonal matrices $X$ in order to determine
the highest $m$. This is now more complicated compared to the case of
$C(\mu,\alpha,\beta)$ due to the
additional generator $R(\nu,\rho,\sigma)$.
Then one determines $F_1$ in the same way as described above.
In order to find the subgroup of diagonal matrices of the form
$\mathrm{diag}(1,x,x^\ast)$, 
one has to multiply the matrices $X$, $EXE^{-1}$ and $BXB^{-1}$ 
with suitable powers of $F_1$. 
Then 
$G$ is the generator of the
resulting cyclic group. 
Finally, $F$ can then be determined by multiplying $F_1$ with an
appropriate power of $G$, as described in the proof of theorem~\ref{abelian}.
\end{appendix}

\end{document}